\documentclass[english]{article}
\pdfoutput=1
\usepackage{lmodern}

\usepackage[T1]{fontenc}
\usepackage[latin9]{inputenc}
\usepackage{booktabs}
\usepackage{mathrsfs}
\usepackage{amsmath}
\usepackage{amssymb}
\usepackage{graphicx}
\usepackage{wasysym}

\makeatletter

\providecommand{\tabularnewline}{\\}

\newcommand{\lyxaddress}[1]{
\par {\raggedright #1
\vspace{1.4em}
\noindent\par}
}

\usepackage{mathrsfs}   
\usepackage{slashed}     
\usepackage{bbold}  
\usepackage{url}
\usepackage{graphicx}
\usepackage[colorlinks=true,linkcolor=redLinks,citecolor=greenLinks,urlcolor=redLinks, pdfborder={0 0 1}]{hyperref}
\usepackage{xcolor}
\usepackage{framed}
\usepackage[numbers,sort&compress]{natbib}
\usepackage{amsmath}

\allowdisplaybreaks

\colorlet{shadecolor}{gray!15}

\definecolor{greenLinks}{rgb}{0, 0.6, 0} 
\definecolor{blueLinks}{rgb}{0, 0, 0.6}
\definecolor{redLinks}{rgb}{0.6, 0, 0}
\definecolor{tempText}{rgb}{0.55, 0.10,0.67}
\definecolor{eprintLinks}{rgb}{0.4, 0.4, 0.4}
\definecolor{journalLinks}{rgb}{0.6, 0, 0}

\newcommand{\MYhref}[3][redLinks]{\href{#2}{\color{#1}{#3}}}%

\usepackage{multirow}
\let\orig@Hy@EveryPageAnchor\Hy@EveryPageAnchor
\def\Hy@EveryPageAnchor{%
    \begingroup
    \hypersetup{pdfview=Fit}%
    \orig@Hy@EveryPageAnchor
    \endgroup
}

\let\oldFootnote\footnote
\newcommand\nextToken\relax

\renewcommand\footnote[1]{%
    \oldFootnote{#1}\futurelet\nextToken\isFootnote}

\newcommand\isFootnote{%
    \ifx\footnote\nextToken\textsuperscript{,}\fi}

\definecolor{myPurple}{RGB}{128,0,182}

\newcommand{\appropto}{\mathrel{\vcenter{
  \offinterlineskip\halign{\hfil$##$\cr
    \propto\cr\noalign{\kern2pt}\sim\cr\noalign{\kern-2pt}}}}}

\makeatother

\usepackage{babel}
\begin{document}

\title{{\Large{}\vspace{-1.0cm}} \hfill {\normalsize{}IFIC/16-33} \\*[10mm] A
flipped 331 model{\Large{}\vspace{0.5cm}}}

\author{{\Large{}Renato M. Fonseca}\thanks{E-mail: renato.fonseca@ific.uv.es} 
\ and {\Large{}Martin Hirsch}\thanks{E-mail: mahirsch@ific.uv.es} 
\date{}}

\maketitle

\lyxaddress{\begin{center}
{\Large{}\vspace{-0.5cm}}AHEP Group, Instituto de Física Corpuscular,
C.S.I.C./Universitat de València\\
Parc Científic de Paterna.  Calle Catedrático José Beltrán, 2 E-46980
Paterna (Valencia) -- Spain
\par\end{center}}


\begin{center}
\today
\par\end{center}
\begin{abstract}
Models based on the extended $SU(3)_{C}\times SU(3)_{L}\times
U(1)_{X}$ (331) gauge group usually follow a common pattern: two
families of left-handed quarks are placed in anti-triplet
representations of the $SU(3)_{L}$ group; the remaining quark family,
as well as the left-handed leptons, are assigned to triplets (or vice-versa). In this
work we present a flipped 331 model where this scheme is reversed:
all three quark families are in the same representation and it is the
lepton families which are discriminated by the gauge symmetry. We
discuss fermion masses and mixing, as well as $Z'$ interactions, in a
minimal model implementing this idea.
\end{abstract}

\section{Introduction}

The idea of embedding the Standard Model (SM) electroweak symmetry
group in $SU(3)_{L}\times U(1)_{X}$ was proposed a long time ago.
Early work, see for example
\cite{Schechter:1974dy,Gupta:1973pv,Albright:1974nd,Georgi:1978bv},
were mostly attempts to explain the absence of flavor changing neutral currents (FCNCs) using an
extended gauge sector instead of a GIM mechanism.  A fully-fledged
three generation model was then proposed by Singer, Valle and
Schechter (SVS) in 1980 \cite{Singer1980}. A ``minimal'' 331 model was
constructed by Pisano, Pleitez \cite{Pisano1992} and Frampton
\cite{Frampton1992} (PPF). Both the SVS and the PPF models, as well as
(nearly) all other 331 models, follow the same, simple construction
principles, as discussed below.

\noindent
Any model beyond the SM needs to satisfy two important
constraints:
\begin{enumerate}
\item Extra fermions should not be introduced, unless
  they are vector-like under the SM group; i.e. they must either be in
  a real representation of $SU(3)_{C}\times SU(2)_{L}\times U(1)_{Y}$,
  or come in pairs of complex conjugated representations.
\item Gauge anomalies should cancel.
\end{enumerate}
These two conditions turn out to be very restrictive for the fermion
sector of 331 models. Just to mention one example, the authors of
\cite{Diaz2005} considered the consequences of gauge anomaly
cancellation (condition 2) in 331 models with $SU(3)_{L}$ singlets,
triplets and anti-triplets. With only those representations the 
standard construction is to place the left-handed lepton doublets of the
SM in three $SU(3)_{L}$ triplets $\psi_{\ell}$ (one per
generation). Then, in order to cancel the $SU(3)_{L}^{3}$ anomaly, for
the left-handed quarks there is no choice except to put two families
in anti-triplets ($Q_{12}$) and the remaining one ($Q_{3}$) in an
$SU(3)_{L}$ triplet (due to the color multiplicity of the quarks). 
Furthermore, one should add $SU(3)_{L}$ singlets for the right-landed
leptons and quarks ($\ell^{c}$, $u^{c}$ and $d^{c}$). The fermion
content of the model is wrapped up by noting that each $SU(3)_{L}$
(anti)triplet contains an $SU(2)_{L}$ doublet plus an extra singlet,
and this latter fermion must be vector-like, hence $SU(3)_{L}$ singlets
with an adequate color and $U(1)_{X}$ charge should be added to the
model ($\ell_{X}$, $J_{12}$ and $J_{3}$). 

Finally, for the symmetry breaking and the generation of fermion
masses, a set of scalars is needed. Again, in the minimal setups,
usually three $SU(3)_{L}$ triplets are introduced to provide the
required Yukawa interactions and gauge boson masses. 

It is important to note that $SU(2)_{L}\times U(1)_{Y}$ can be embedded
in $SU(3)_{L}\times U(1)_{X}$ in more than one way; this can
be parametrized by a number $\beta$ which relates the hypercharge
$Y$, the $T_{8}$ generator of $SU(3)_{L}$ and $X$ as follows:
\begin{align}
Y & =\beta T_{8}+X\,.
\end{align}
In fact, it turns out that one can build minimal models obeying the
two conditions above for an arbitrary $\beta$ with the representations
in table \ref{tab:Fields_of_standard_331_models}. 
\footnote{Throughout this
  work $SU(2)_{L}$ representations have hats to distinguish them from
  $SU(3)_{L}$ representations.} 

\begin{table}[tbph]
\begin{centering}
\begin{tabular}{cccc}
\toprule 
Name & 331 representation & SM group decomposition & \# flavors\tabularnewline
\midrule
$\psi_{\ell}$ & $\left(\mathbf{1},\mathbf{3},-\frac{1}{2}-\frac{1}{2\sqrt{3}}\beta\right)$ & $\left(\mathbf{1},\widehat{\mathbf{2}},-\frac{1}{2}\right)+\left(\mathbf{1},\widehat{\mathbf{1}},-\frac{1}{2}-\frac{\sqrt{3}}{2}\beta\right)$ & 3\tabularnewline
$\ell^{c}$ & $\left(\mathbf{1},\mathbf{1},1\right)$ & $\left(\mathbf{1},\widehat{\mathbf{1}},1\right)$ & 3\tabularnewline
$\ell_{X}$ & $\left(\mathbf{1},\mathbf{1},\frac{1}{2}+\frac{\sqrt{3}}{2}\beta\right)$ & $\left(\mathbf{1},\widehat{\mathbf{1}},\frac{1}{2}+\frac{\sqrt{3}}{2}\beta\right)$ & 3\tabularnewline
$Q_{12}$ & $\left(\mathbf{3},\overline{\mathbf{3}},\frac{1}{6}+\frac{1}{2\sqrt{3}}\beta\right)$ & $\left(\mathbf{3},\widehat{\mathbf{2}},\frac{1}{6}\right)+\left(\mathbf{3},\widehat{\mathbf{1}},\frac{1}{6}+\frac{\sqrt{3}}{2}\beta\right)$ & 2\tabularnewline
$Q_{3}$ & $\left(\mathbf{3},\mathbf{3},\frac{1}{6}-\frac{1}{2\sqrt{3}}\beta\right)$ & $\left(\mathbf{3},\widehat{\mathbf{2}},\frac{1}{6}\right)+\left(\mathbf{3},\widehat{\mathbf{1}},\frac{1}{6}-\frac{\sqrt{3}}{2}\beta\right)$ & 1\tabularnewline
$u^{c}$ & $\left(\overline{\mathbf{3}},\mathbf{1},-\frac{2}{3}\right)$ & $\left(\overline{\mathbf{3}},\widehat{\mathbf{1}},-\frac{2}{3}\right)$ & 3\tabularnewline
$d^{c}$ & $\left(\overline{\mathbf{3}},\mathbf{1},\frac{1}{3}\right)$ & $\left(\overline{\mathbf{3}},\widehat{\mathbf{1}},\frac{1}{3}\right)$ & 3\tabularnewline
$J_{12}$ & $\left(\overline{\mathbf{3}},\mathbf{1},-\frac{1}{6}-\frac{\sqrt{3}}{2}\beta\right)$ & $\left(\overline{\mathbf{3}},\widehat{\mathbf{1}},-\frac{1}{6}-\frac{\sqrt{3}}{2}\beta\right)$ & 2\tabularnewline
$J_{3}$ & $\left(\overline{\mathbf{3}},\mathbf{1},-\frac{1}{6}+\frac{\sqrt{3}}{2}\beta\right)$ & $\left(\overline{\mathbf{3}},\widehat{\mathbf{1}},-\frac{1}{6}+\frac{\sqrt{3}}{2}\beta\right)$ & 1\tabularnewline
\midrule
$\phi_{1}$ & $\left(\mathbf{1},\mathbf{3},\frac{1}{2}-\frac{1}{2\sqrt{3}}\beta\right)$ & $\left(\mathbf{1},\widehat{\mathbf{2}},\frac{1}{2}\right)+\left(\mathbf{1},\widehat{\mathbf{1}},\frac{1}{2}-\frac{\sqrt{3}}{2}\beta\right)$ & 1\tabularnewline
$\phi_{2}$ & $\left(\mathbf{1},\mathbf{3},\frac{1}{\sqrt{3}}\beta\right)$ & $\left(\mathbf{1},\widehat{\mathbf{2}},\frac{\sqrt{3}}{2}\beta\right)+\left(\mathbf{1},\widehat{\mathbf{1}},0\right)$ & 1\tabularnewline
$\phi_{3}$ & $\left(\mathbf{1},\mathbf{3},-\frac{1}{2}-\frac{1}{2\sqrt{3}}\beta\right)$ & $\left(\mathbf{1},\widehat{\mathbf{2}},-\frac{1}{2}\right)+\left(\mathbf{1},\widehat{\mathbf{1}},-\frac{1}{2}-\frac{\sqrt{3}}{2}\beta\right)$ & 1\tabularnewline
\bottomrule
\end{tabular}
\par\end{centering}

\protect\caption{\label{tab:Fields_of_standard_331_models}Representations
  used in a ``standard'' 331 model for a generic $\beta$ parameter. The
  $\phi_{i}$ are scalars while all other fields are left-handed Weyl
  fermions.  It is also possible to flip the sign of $\beta$ and at
  the same time swap all $SU(3)_{L}$ representations to the
  corresponding anti-representations.}
\end{table}

For example, the Singer-Valle-Schechter model \cite{Singer1980}
is obtained by substituting $\beta=-1/\sqrt{3}$, while the
Pisano-Pleitez-Frampton model \cite{Pisano1992,Frampton1992}
uses $\beta=-\sqrt{3}$. The various 331 models may also differ from
one-another in the scalar sector or perhaps due to the addition of
vector fermions (under the full 331 group) --- indeed, such modifications
might be needed to account for the observed
charged lepton \cite{Foot1993} and neutrino masses \cite{Fonseca2016}.
Furthermore, for some
values of $\beta$, some of the representations in the table might be
vector-like (or form vector-like pairs) hence they might be removed
without violating the conditions 1 and 2 in the text. This reduces the
number of fermion fields in the SVS model to 18 and in the PPF version
to 15.

Switching triplets with anti-triplets of $SU(3)_{L}$ and changing
$\beta$ to $-\beta$ is also allowed and this leads to pairs of
models, with rather similar phenomenology but subtly different FCNC
effects, suppressed by $(\Lambda_{EW}/\Lambda_{331})^2$ \cite{Buras:2014yna}. Ignoring running effects, it is easy to show that the gauge
couplings of $SU(3)_{L}/SU(2)_{L}$, $U(1)_{Y}$ and $U(1)_{X}$ obey the
relation $g_{Y}^{-2}=g_{X}^{-2}+\beta^{2}g_{L}^{-2}$ therefore the
requirement that $g_{X}^{2}$ is positive implies that
$\left|\beta\right|\leq g_{L}/g_{Y}=\tan^{-1}\theta_{W}\approx1.8$.
If we add the condition that no fractionally charged leptons should
appear in the physical spectrum, \footnote{Fractionally charged
  leptons, produced abundantly in the early universe, would lead to an
  unacceptable cosmology, because there are no decay modes for the
  lightest of these exotic states.} only four values of $\beta$ lead to
viable models: $\beta = \pm \sqrt{3},\pm 1/\sqrt{3}$. With the
above-mentioned switch to anti-representations this makes a total
of 8 possible basic models.

Finally we mention the 331 model which is obtainable from an $E_{6}$
grand unified theory \cite{Sanchez2001}. This model is a notable
exception to the scheme of table
\ref{tab:Fields_of_standard_331_models} because in it all three
families of quarks and leptons are in equal $SU(3)_{C}\times
SU(3)_{L}\times U(1)_{X}$ representations. It is curious to note 
that in this model anomaly cancellation occurs within one family
--- as in the Standard Model --- thus dispelling 
the notion that 331 models predict the number of families to be 
equal to the number of colours. 

\section{A flipped 331 model}

As we have mentioned already, in addition to choosing different values of $\beta$, some simple
adjustments can be made to the class of 331 models mentioned so far:
(a) the scalar sector may be changed, for example by adding more
fields or (b) fermions which are vector-like under the full 331
symmetry might be added to the model. 
More interesting, however, is the observation that gauge anomaly
cancellation does not force 331 models to follow the structure of table
\ref{tab:Fields_of_standard_331_models}.  Indeed, the field content in
table \ref{tab:Fields_of_new_model} is not only anomaly free; it has
in fact the same total number of fermion representations as the
standard 331 model with the same $\beta$. Since in this model there is perfect 
quark family replication, while leptons are 
placed into different representations, we call this setup a flipped 331 
model.

\begin{table}[tbph]
\begin{centering}
\begin{tabular}{ccccc}
\toprule 
Name & 331 rep. & SM group decomposition & Components & \# flavors\tabularnewline
\midrule
$L_{e}$ & $\left(\mathbf{1},\mathbf{6},-\frac{1}{3}\right)$ & $\left(\mathbf{1},\widehat{\mathbf{3}},0\right)+\left(\mathbf{1},\widehat{\mathbf{2}},-\frac{1}{2}\right)+\left(\mathbf{1},\widehat{\mathbf{1}},-1\right)$ & $\left(\begin{array}{ccc}
\Sigma^{+} & \frac{1}{\sqrt{2}}\Sigma^{0} & \frac{1}{\sqrt{2}}\nu_{e}\\
\frac{1}{\sqrt{2}}\Sigma^{0} & \Sigma^{-} & \frac{1}{\sqrt{2}}\ell_{e}\\
\frac{1}{\sqrt{2}}\nu_{e} & \frac{1}{\sqrt{2}}\ell_{e} & E_{e}
\end{array}\right)$ & 1\tabularnewline
$L_{\alpha=\mu,\tau}$ & $\left(\mathbf{1},\mathbf{3},-\frac{2}{3}\right)$ & $\left(\mathbf{1},\widehat{\mathbf{2}},-\frac{1}{2}\right)+\left(\mathbf{1},\widehat{\mathbf{1}},-1\right)$ & $\left(\nu_{\alpha},\ell_{\alpha},E_{\alpha}\right)^{T}$ & 2\tabularnewline
$\ell_{\alpha}^{c}$ & $\left(\mathbf{1},\mathbf{1},1\right)$ & $\left(\mathbf{1},\widehat{\mathbf{1}},1\right)$ & $\ell_{\alpha}^{c}$ & 6\tabularnewline
$Q_{\alpha}$ & $\left(\mathbf{3},\overline{\mathbf{3}},\frac{1}{3}\right)$ & $\left(\mathbf{3},\widehat{\mathbf{2}},\frac{1}{6}\right)+\left(\mathbf{3},\widehat{\mathbf{1}},\frac{2}{3}\right)$ & $\left(d_{\alpha},-u_{\alpha},U_{\alpha}\right)^{T}$ & 3\tabularnewline
$u_{\alpha}^{c}$ & $\left(\overline{\mathbf{3}},\mathbf{1},-\frac{2}{3}\right)$ & $\left(\overline{\mathbf{3}},\widehat{\mathbf{1}},-\frac{2}{3}\right)$ & $u_{\alpha}^{c}$ & 6\tabularnewline
$d_{\alpha}^{c}$ & $\left(\overline{\mathbf{3}},\mathbf{1},\frac{1}{3}\right)$ & $\left(\overline{\mathbf{3}},\widehat{\mathbf{1}},\frac{1}{3}\right)$ & $d_{\alpha}^{c}$ & 3\tabularnewline
\midrule
$\phi_{i=1,2}$ & $\left(\mathbf{1},\mathbf{3},\frac{1}{3}\right)$ & $\left(\mathbf{1},\widehat{\mathbf{2}},\frac{1}{2}\right)+\left(\mathbf{1},\widehat{\mathbf{1}},0\right)$ & $\left(H_{i}^{+},H_{i}^{0},\sigma_{i}^{0}\right)^{T}$ & 2\tabularnewline
$\phi_{3}$ & $\left(\mathbf{1},\mathbf{3},-\frac{2}{3}\right)$ & $\left(\mathbf{1},\widehat{\mathbf{2}},-\frac{1}{2}\right)+\left(\mathbf{1},\widehat{\mathbf{1}},-1\right)$ & $\left(H_{3}^{0},H_{3}^{-},\sigma_{3}^{-}\right)^{T}$ & 1\tabularnewline
$S$ & $\left(\mathbf{1},\mathbf{6},\frac{2}{3}\right)$ & $\left(\mathbf{1},\widehat{\mathbf{3}},1\right)+\left(\mathbf{1},\widehat{\mathbf{2}},\frac{1}{2}\right)+\left(\mathbf{1},\widehat{\mathbf{1}},0\right)$ & $\left(\begin{array}{ccc}
\Delta^{++} & \frac{1}{\sqrt{2}}\Delta^{+} & \frac{1}{\sqrt{2}}H_{S}^{+}\\
\frac{1}{\sqrt{2}}\Delta^{+} & \Delta^{0} & \frac{1}{\sqrt{2}}H_{S}^{0}\\
\frac{1}{\sqrt{2}}H_{S}^{+} & \frac{1}{\sqrt{2}}H_{S}^{0} & \sigma_{S}^{0}
\end{array}\right)$ & 1\tabularnewline
\bottomrule
\end{tabular}
\par\end{centering}

\protect\caption{\label{tab:Fields_of_new_model}Representations for the
  flipped 331 model. The $\phi$'s and $S$ are scalars while all other
  fields are left-handed Weyl fermions. The letters $i$ and $\alpha$
  stand for flavor indices, going from 1 to the number quoted in the
  last column (except for $L_{\alpha=\mu,\tau}$).  The components of
  the $SU(3)_{L}$ triplets in the forth column match the ordering of
  the SM group decomposition given in the second column; as for the
  sextets, $\Sigma$ and $\Delta$ stand for the triplet
  sub-representations, $\left(\nu_{e},\ell_{e}\right)^{T}$ and
  $\left(H_{S}^{+},H_{S}^{0}\right)^{T}$ form $SU(2)_{L}$ doublets,
  while $E_{e}$ and $\sigma_{S}^{0}$ are singlets.}
\end{table}

Anomaly cancellation works in the flipped 331 model as follows: an
$SU(3)_{L}$ sextet contributes as much as 7 triplets to the
$SU(3)_{L}^{3}$ anomaly, which at first seems problematic. Not only is
this number large, it is not directly relatable to the number of
families (3) nor the number of colors (also 3). The simplest solution to such a large
anomaly contribution is to make the sextet field a lepton; otherwise, if it were a quark,
the color multiplicity would compound the problem of
canceling the $SU(3)_{L}^{3}$ anomaly. The decomposition of a sextet
representation is as follows (for a generic $\beta$):
\begin{align}
\left(\mathbf{1},\mathbf{6},x\right) & \rightarrow\left(\mathbf{1},\widehat{\mathbf{3}},x+\frac{1}{\sqrt{3}}\beta\right)+\left(\mathbf{1},\widehat{\mathbf{2}},x-\frac{1}{2\sqrt{3}}\beta\right)+\left(\mathbf{1},\widehat{\mathbf{1}},x-\frac{2}{\sqrt{3}}\beta\right)\,.\label{eq:6decomposition}
\end{align}
Significantly, the sextet contains an $SU(2)_{L}$ triplet. This
triplet will have necessarily an electroweak scale mass, unless it is
made a vector particle under the SM gauge group.  To do that one could
introduce other (bigger) $SU(3)_{L}$ multiplets to find a
$\left(\mathbf{1},\widehat{\mathbf{3}},-x-\frac{1}{\sqrt{3}}\beta\right)$
state which can form a vector pair with the triplet in equation
(\ref{eq:6decomposition}). While such a construction might indeed be
possible, it runs the risk of requiring a large number of extra
representations (see section 3.6 of \cite{Fonseca2015}).  Yet, there
is a simpler alternative. For $x=-\frac{1}{\sqrt{3}}\beta$ the triplet
in equation (\ref{eq:6decomposition}) becomes a real representation,
thereby solving the problem. Note that, if the sextet was a quark,
this idea would not work.

We will thus focus on the possibility of identifying the doublet
inside the sextet as one of the families of the left-handed
leptons:
$\widehat{L}_{e}=\left(\mathbf{1},\widehat{\mathbf{2}},-1/2\right)$.
(The reason for associating it with the electron will be made clear in
the next section.)  In that case,
$x-\frac{1}{2\sqrt{3}}\beta=-\frac{\sqrt{3}}{2}\beta$ should equate to
$-1/2$, hence
\begin{align}
\beta & =\frac{1}{\sqrt{3}}\,.
\end{align}
The two remaining families of left-handed leptons, $\widehat{L}_{\mu}$
and $\widehat{L}_{\tau}$, can then be put into either triplets or
anti-triplets of $SU(3)_{L}$. However, the triplet possibility is clearly
more interesting for the following reason: one sextet and two triplets 
provide an anomaly contribution exactly equal to the one of nine triplets.  
Hence, if all three left-handed quarks are placed
in anti-triplets of $SU(3)_{L}$, the $SU(3)_{L}^{3}$ cancels. 
It is then rather simple to add the correct $SU(3)_{L}$ singlets in
order to reproduce the SM chirality (condition 1 above). And,
surprisingly, without further tuning the model, it turns out that not
only does the $SU(3)_{L}^{3}$ gauge anomaly cancel, but in fact all
gauge anomalies cancel in this minimal setup (see table \ref{tab:Anomalies}). Thus, no additional
particles are needed in the fermion sector of this minimal flipped
model, which is summarized in table \ref{tab:Fields_of_new_model}.

With conditions 1 and 2 satisfied, all that remains is to find
adequate scalar fields. Using the notation in table
\ref{tab:Fields_of_new_model}, one should have at least the Yukawa
interactions $Qu^{c}\times\textrm{scalar}$, $Qd^{c}\times\textrm{scalar}$,
$L_{\mu\tau}\ell^{c}\times\textrm{scalar}$, $L_{e}L_{e}\times\textrm{scalar}$ 
and $L_{e}\ell^{c}\times\textrm{scalar}$.  The minimal setup to do this
contains three scalar triplets $\phi_{1,2,3}$ and a scalar sextet
$S$. Note that two copies of the same triplet
representation, $\phi_{1}$ and $\phi_{2}$, are needed for fermion
masses, as explained below. 

\begin{center}
	\begin{table}[tbh]
		\begin{centering}
			\def\arraystretch{1.3}%
						\begin{tabular}{ccccccc}
				\toprule 
				Field & $SU(3)_{C}^{3}$ & $SU(3)_{L}^{3}$ & $SU(3)_{C}^{2}U(1)_{X}$ & $SU(3)_{L}^{2}U(1)_{X}$ & $U(1)_{X}^{3}$ & \# flavors\tabularnewline
				\midrule
				$L_{e}$ & $\phantom{-}0$ & $\phantom{-}\frac{7}{2}$ & $\phantom{-}0$ & $-\frac{5}{6}$ & $-\frac{2}{9}$ & 1\tabularnewline
				$L_{\alpha=\mu,\tau}$ & $\phantom{-}0$ & $\phantom{-}\frac{1}{2}$ & $\phantom{-}0$ & $-\frac{1}{3}$ & $-\frac{8}{9}$ & 2\tabularnewline
				$\ell_{\alpha}^{c}$ & $\phantom{-}0$ & $\phantom{-}0$ & $\phantom{-}0$ & $\phantom{-}0$ & $\phantom{-}1$ & 6\tabularnewline
				$Q_{\alpha}$ & $\phantom{-}\frac{3}{2}$ & $-\frac{3}{2}$ & $\phantom{-}\frac{1}{2}$ & $\phantom{-}\frac{1}{2}$ & $\phantom{-}\frac{1}{3}$ & 3\tabularnewline
				$u_{\alpha}^{c}$ & $-\frac{1}{2}$ & $\phantom{-}0$ & $-\frac{1}{3}$ & $\phantom{-}0$ & $-\frac{8}{9}$ & 6\tabularnewline
				$d_{\alpha}^{c}$ & $-\frac{1}{2}$ & $\phantom{-}0$ & $\phantom{-}\frac{1}{6}$ & $\phantom{-}0$ & $\phantom{-}\frac{1}{9}$ & 3\tabularnewline
				\bottomrule
			\end{tabular}\def\arraystretch{1}
			\par\end{centering}
		
		\protect\caption{\label{tab:Anomalies}Contribution of one copy/flavor of a given fermion
			representation to each of the five gauge anomalies. The sum of all
			rows, weighted by the number of flavors in the last column, adds up
			to zero, hence the model is anomaly free.}		
	\end{table}
	
	\par\end{center}

\section{Fermion masses and FCNC constraints}

We shall not consider in detail the scalar potential of the model. 
Instead we assume that there exists a stable, charge-preserving vacuum
state and study the consequences. In this spirit, we allow all neutral
scalar components to have a non-zero vacuum expectation value (VEV):
\begin{gather}
\left\langle \sigma_{1,2}^{0}\right\rangle \equiv n_{1,2},\;\left\langle H_{1,2,3}^{0}\right\rangle \equiv k_{1,2,3},\;\left\langle \sigma_{S}^{0}\right\rangle \equiv n_{S},\;\left\langle H_{S}^{0}\right\rangle \equiv k_{S},\;\left\langle \Delta^{0}\right\rangle \equiv\epsilon_{S}\,.
\end{gather}
In this notation, all $n_{\alpha}$ stand for VEVs which are singlets
of the SM gauge group, all $k_{\alpha}$ are associated with SM
doublets, and $\epsilon_{S}$ is the VEV of the SM triplet. Hence, in
principle one expects that $\epsilon_{S}\ll k_{\alpha}\ll
n_{\alpha}$.

The following Yukawa interactions are allowed by the gauge symmetry:
\begin{align}
\mathscr{L}_{\textrm{Yukawa}} & =\mathscr{L}_{\textrm{leptons}}+\mathscr{L}_{\textrm{quarks}}\,,\\
\mathscr{L}_{\textrm{leptons}} & =\sum_{i=1}^{2}y_{\alpha\beta}^{\ell(i)}L_{\alpha}\ell_{\beta}^{c}\phi_{i}^{*}+y_{\alpha}^{\ell\prime}L_{e}\ell_{\alpha}^{c}S^{*}+y''L_{e}L_{e}S+\textrm{h.c.}\,,\\
\mathscr{L}_{\textrm{quarks}} & =\sum_{i=1}^{2}y_{\alpha\beta}^{u(i)}Q_{\alpha}u_{\beta}^{c}\phi_{i}+y_{\alpha\beta}^{d}Q_{\alpha}d_{\beta}^{c}\phi_{3}+\textrm{h.c.}\,.
\end{align}
Here, repeated Greek indices $\alpha,\beta$, representing flavors of
the fermion fields, are assumed to be summed over.  Fermion masses are
generated from these interactions once the scalar fields acquire VEVs,
\begin{align*}
\mathscr{L}_{\textrm{fermion mass}} & =m_{\alpha\beta}^{\ell}\Psi_{\alpha}^{\ell}\Psi_{\beta}^{\ell^{c}}+m_{\alpha\beta}^{\nu}\Psi_{\alpha}^{\nu}\Psi_{\beta}^{\nu}+m_{\alpha\beta}^{u}\Psi_{\alpha}^{u}\Psi_{\beta}^{u^{c}}+m_{\alpha\beta}^{d}\Psi_{\alpha}^{d}\Psi_{\beta}^{d^{c}}+\textrm{h.c.}\,.
\end{align*}

The quark sector of the model is rather simple. Quark masses
can be accommodated at tree level with a suitable choice of VEVs and
Yukawa couplings $y^{u(1)}$, $y^{u(2)}$, $y^{d}$ since
\begin{align}
	m^{u}= &\; \left(\begin{array}{c}
		y_{\alpha\beta}^{u(1)}k_{1}+y_{\alpha\beta}^{u(2)}k_{2}\\
		y_{\alpha\beta}^{u(1)}n_{1}+y_{\alpha\beta}^{u(2)}n_{2}
	\end{array}\right)\,,\label{eq:mup}\\
	m^{d}= &\; y_{\alpha\beta}^{d} k_{3}\,,\label{eq:mdown}
\end{align}
in the basis $\Psi^{u}=\left(u_{\alpha},U_{\alpha}\right)^{T}$,
$\Psi^{u^{c}}=\left(u_{\beta}^{c}\right)$,
$\Psi^{d}=\left(d_{\alpha}\right)$,
$\Psi^{d^{c}}=\left(d_{\beta}^{c}\right)$. From equation \eqref{eq:mup} it
is straightforward to understand why the model needs two copies of
$\phi_{1,2}$: in the case of only one $\phi$, the mass matrix $m^u$
has rank 3, thus generating three massless up-quarks. On the other
hand, with two copies of $\phi$ and requiring $k_1/n_1 \ne k_2/n_2$,
the model is able to fit the data, but there are no predictions in the
quark sector.

Let us now consider the charged lepton mass matrix in the basis
$\Psi^{\ell}=\left(\ell_{\alpha},E_{\alpha},E_{e},\ell_{e},\Sigma^{-}\right)^{T}$
and $\Psi^{\ell^{c}}=\left(\ell_{\beta}^{c},\Sigma^{+}\right)^{T}$
(where $\alpha=\mu,\tau$; $\beta=1,\cdots,6$):
\begin{align}
m^{\ell} & =\left(\begin{array}{cc}
y_{\alpha\beta}^{\ell(1)}k_{1}+y_{\alpha\beta}^{\ell(2)}k_{2} & 0\\
y_{\alpha\beta}^{\ell(1)}n_{1}+y_{\alpha\beta}^{\ell(2)}n_{2} & 0\\
n_{S}y'_{\beta} & -\epsilon_{S}y''\\
k_{S}y'_{\beta} & k_{S}y''\\
\epsilon_{S}y'_{\beta} & -n_{S}y''
\end{array}\right)\,,
\end{align}
A striking feature of this matrix is that there is a combination of
the last three rows which adds up to 0, hence there is a massless
combination of the $\Psi_{\alpha}^{\ell}$ and of the
$\Psi_{\alpha}^{\ell^{c}}$ as well. The latter turns out to be a
combination of the six $\ell_{\beta}^{c}$ so it is a pure
$\left(\mathbf{1},\widehat{\mathbf{1}},1\right)$ state under the
Standard Model group. As for the negatively charged massless state, it
is easy to check that it corresponds to $\propto
k_{S}\Sigma^{-}+\left(n_{S}-\epsilon_{S}\right)\ell_{e}-k_{S}E_{e}$.
A natural possibility, although perhaps not the only one, is to
associate this combination to the left-handed electron, the lightest
of the charged leptons --- hence the subscript $e$ in $L_{e}$. The
admixture of $\Sigma^{-}$ (part of an $SU(2)_{L}$ triplet) and of
$E_{e}$ (an $SU(2)_{L}$ singlet) is suppressed by the ratio
$k_{S}/n_{S}$. This ratio has to be small, due to the smallness 
of neutrino masses, see below. Note also that in the limit 
$y'' \to 0$ there appears a second massless eigenstate. This 
limit is of course unphysical, since it corresponds to a 
fourth light charged lepton.

\begin{figure}[tbph]
\begin{centering}
\includegraphics[scale=0.85]{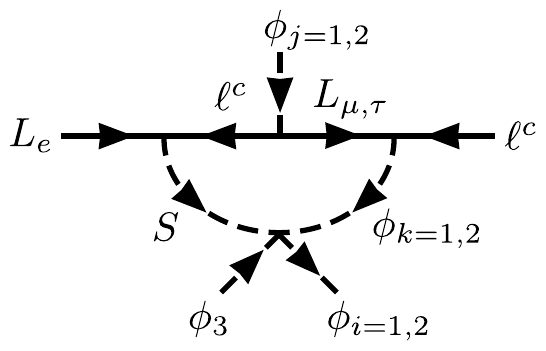}
\par\end{centering}

\protect\caption{\label{fig:electron_mass}One-loop diagram responsible
  for the generation of a small electron mass.}
\end{figure}

The electron mass can be generated by radiative corrections to the
matrix $m^{\ell}$ --- specifically to the block dependent on $y'$.
For example, the effective operator 
$\mathcal{O}_{\alpha}^{(ij)}L_{e}\ell_{\alpha}^{c}\phi_{i}^{*}\phi_{j}\phi_{3}$
($i,j=1,2$), see figure \ref{fig:electron_mass}, generates a 
contribution, roughly of the order of:
\begin{align}
\textrm{Fig. 1 diagram} & \propto\sum_{j=1}^{2}\frac{\lambda_{ik}k_{3}}{16\pi^{2}\Lambda^{2}}y'_{\gamma}y_{\delta\gamma}^{\ell(j)*}y_{\delta\alpha}^{\ell(k)}\ell_{\alpha}^{c}\left(-k_{j}n_{i}E_{e}+\frac{n_{i}n_{j}-k_{i}k_{j}}{\sqrt{2}}\ell_{e}+k_{i}n_{j}\Sigma^{-}\right)\, .
\end{align}
Here, $\Lambda$ is of the order of the 331 breaking scale and
$\lambda_{ij}$ is the coupling constant of the quartic scalar
interaction $S\phi_{3}\phi_{i}^{*}\phi_{j}$ ($i,j=1,2)$.  Inserting
very roughly $n_i \sim \Lambda \sim 10^3$ GeV, $k_3 \sim 100$ GeV and all
couplings $y \sim y' \sim \lambda_{ij} \sim 0.2$ results in a mass
correction of the MeV order.

We now turn to a discussion of neutrino masses. The fermion content 
of the model contains four colorless neutral fields. With the ordering
$\Psi^{\nu}=\left(\nu_{\mu},\nu_{\tau},\nu_{e},\Sigma^{0}\right)^{T}$
the tree-level mass matrix reads
\begin{align}
m^{\nu} & =y''\left(\begin{array}{cccc}
0 & 0 & 0 & 0\\
0 & 0 & 0 & 0\\
0 & 0 & \epsilon_{S} & -\frac{k_{S}}{\sqrt{2}}\\
0 & 0 & -\frac{k_{S}}{\sqrt{2}} & n_{S}
\end{array}\right)\, ,\label{eq:mNu}
\end{align}
i.e. at tree-level there are two massless neutrinos, $\nu_{\mu}$ and
$\nu_{\tau}$. In the seesaw approximation, the two non-zero eigenstates have
 masses $m_{\Sigma} \simeq y''n_{S}$ and $m_{\nu}
\simeq
y''\left[\epsilon_{S}-k_{S}^{2}/\left(2n_{S}\right)\right]$. The
latter is a mixture of a type-II and type-III seesaw mechanism
contributions. Unless one considers the extremely fine-tuned situation
where $\epsilon_{S} \equiv k_{S}^{2}/\left(2n_{S}\right)$, this
implies that $k_{S}$ has to much smaller than all other $k_i$,
i.e. for $n_S \sim$ TeV one needs $k_S$ to be below $10^{-3}$ GeV.

\begin{figure}[tbph]
\begin{centering}
\includegraphics[scale=0.85]{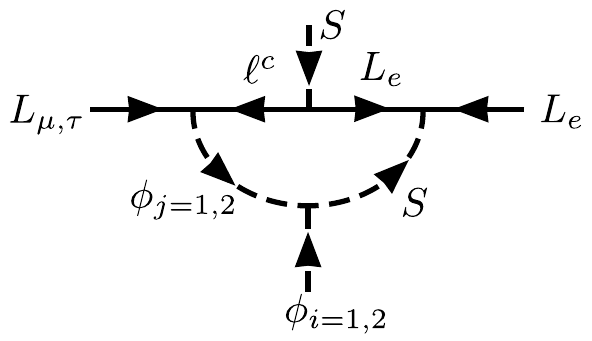}
\par\end{centering}

\protect\caption{\label{fig:neutrino_mass}Loop diagram responsible for
  the generation of two of the light neutrino masses as well as the
  mixing of $\nu_e$ with the other two light neutrinos.}
\end{figure}

This tree-level picture is clearly insufficient, since oscillation
data requires (a) that at least two neutrinos should be massive, and (b)
that $\nu_{e}$ mixes with the two other light
neutrinos. We thus have also to consider loops contributing to the
neutrino mass matrix. We are especially interested in effective
operators of the type $\mathcal{O}_{\alpha}^{\prime(i)}/\Lambda
L_{\alpha}L_{e}\phi_{i}S$ ($\alpha=\mu,\tau$; $i=1,2$), since these 
will generate contributions mixing $\nu_{e}$ with $\nu_{\mu}$ and 
$\nu_{\tau}$. Indeed, as shown in figure
\ref{fig:neutrino_mass}, such type of operator is already present 
in the model, using only the minimal set of scalars given 
in table \ref{tab:Fields_of_new_model}. A rough estimate of this 
diagram results in
\begin{eqnarray}\label{eq:mnulp}
\textrm{Fig. 2 diagram} & \propto & \sum_{j=1}^{2}
\frac{f_{ij}}{16\pi^{2}\Lambda^{2}}y_{\alpha\beta}^{\ell(j)}y_{\beta}^{\ell\prime*}y''\nu_{\alpha}\left[\left(-\frac{k_{i}k_{S}}{\sqrt{2}}+n_{i}\epsilon_{S}\right)\nu_{e}+\left(-\frac{n_{i}k_{S}}{\sqrt{2}}+k_{i}n_{S}\right)\Sigma^{0}\right]\,,\\
\nonumber
 & \equiv & \omega_{e\mu}\nu_{e}\nu_{\mu}+\omega_{e\tau}\nu_{e}\nu_{\tau}+\omega_{\Sigma\mu}\nu_{\Sigma}\nu_{\mu}+\omega_{\Sigma\tau}\nu_{\Sigma}\nu_{\tau}\,,
\end{eqnarray}
where $f_{ij}=f_{ji}$ represents the coupling constant of the
trilinear scalar interaction $\phi_{i}\phi_{j}S$ ($i,j=1,2)$. Note
that equation  \eqref{eq:mnulp} does not include the corresponding loop
functions and thus serves only as a rough order-of-magnitude estimate.

There is sufficient freedom in the Yukawa couplings and the VEVs to
consider the four $\omega_{\alpha\beta}$ coefficients to be essentially
free parameters.  Inserting the $\omega_{\alpha\beta}$
coefficients in the matrix $m^{\nu}$ shown in equation (\ref{eq:mNu}),
and integrating out the heavy $\Sigma^{0}$ state with the standard
seesaw formula, we obtain
\begin{align}
m_{\textrm{light}}^{\nu} & \approx\left(\begin{array}{ccc}
\epsilon_{S}-\frac{k_{S}^{2}}{2n_{S}} & \omega_{e\mu}+\omega_{\Sigma\mu}\frac{k_{S}}{\sqrt{2}n_{S}} & \omega_{e\tau}+\omega_{\Sigma\tau}\frac{k_{S}}{\sqrt{2}n_{S}}\\
\omega_{e\mu}+\omega_{\Sigma\mu}\frac{k_{S}}{\sqrt{2}n_{S}} & -\frac{\omega_{\Sigma\mu}^{2}}{n_{S}} & -\frac{\omega_{\Sigma\mu}\omega_{\Sigma\tau}}{n_{S}}\\
\omega_{e\tau}+\omega_{\Sigma\tau}\frac{k_{S}}{\sqrt{2}n_{S}} & -\frac{\omega_{\Sigma\mu}\omega_{\Sigma\tau}}{n_{S}} & -\frac{\omega_{\Sigma\tau}^{2}}{n_{S}}
\end{array}\right)
\end{align}
in the $\left(\nu_{e},\nu_{\mu},\nu_{\tau}\right)$ basis, to a good
approximation (for simplicity, we ignore here the charged lepton rotation matrix, which at tree level only affects the $\mu\tau$ sector). Notice that the structure in the $\mu\tau$ block of
this matrix has determinant zero. Neutrino oscillation data
\cite{Forero2014} can be fitted to a matrix with such a property, 
but the fit strongly prefers inverse hierarchy with a small 
but slightly non-zero value for $m_{\nu_3}$. One can also easily 
check that for $y_{\alpha\beta}^{\ell(j)} \sim y_{\beta}^{\ell\prime} \sim 
{\cal O}(10^{-1})$,  $y'' \sim {\cal O}(1)$, the scales $k_i \sim 200$ GeV,  
$n_S \sim \Lambda \sim {\cal O}({\rm TeV})$ and $k_S \sim 10^{-4}$ 
GeV, one requires roughly $f_{ij} \sim 10$ GeV in order to get entries 
in the neutrino mass matrix of order few $10$'s of meV, the magnitude 
required by the experimental data for inverse hierarchical neutrinos. 

Thus, the minimal model, as defined in table \ref{tab:Fields_of_new_model},
predicts an inverse hierarchy for neutrinos. However, it
should be kept in mind that with more scalars added to the model,
other fits to neutrino data will become possible as well.

Finally, we turn our attention to flavor changing neutral current
effects in the quark and charged lepton sectors. In addition to the
Standard Model photon and $Z$ boson, 331 models contain a $Z'$ boson
with fermion interactions which are essentially diagonal in the
flavor basis. In standard 331 models, since all three lepton families
are in the same group representation, there are no $Z'$ lepton
flavor changing interactions. On the other hand, left-handed quarks
are placed in one triplet plus two anti-triplet representations of
$SU(3)_{L}$. For the $Z'$ quark-antiquark interactions there appears
then a matrix $\eta=\textrm{diag}\left(1,-1,-1\right)$ in the flavor
basis which, once the quarks are rotated to the mass eigenstate basis,
produces a vertex proportional to the combination $V^{\dagger}\eta V$, where $V$ is the
rotation matrix for left-handed quarks. Thus, the GIM mechanism 
does not work in $Z'$ interactions with quarks in standard 331 models. 
This results in stringent lower limits on the $Z'$ mass \cite{Buras:2015kwd}.

\begin{figure}[tbph]
\begin{centering}
\includegraphics[scale=0.85]{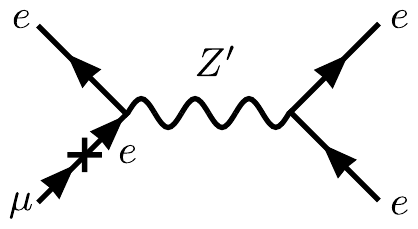}
\par\end{centering}

\protect\caption{\label{fig:muToeee}The $\mu\rightarrow
  e\overline{e}e$ decay mediated by the $Z'$ gauge boson in the
  flipped 331 model.}
\end{figure}

In our model, the situation is reversed. Now all left-handed quark families are in
anti-triplet representations, thus the only limits on $Z'$ from FCNC
observables come from terms mixing ordinary with exotic quarks.  
Note that this mixing can be made arbitrarily small \cite{Boucenna2015b}. 
For this reason, anomalies in rare kaon an $B$ decays cannot be explained by
$Z'$ interactions in our setup.
However, since leptons are in different multiplets in our model, now
potentially dangerous $Z'$ interactions will appear in the lepton
sector (see figure \ref{fig:muToeee}). In the flavor basis, the
relevant Lagrangian term is
\begin{align}
\mathscr{L}_{\ell Z'} & =i\left(\sqrt{3g_{L}^{2}-g_{Y}^{2}}y_{\alpha}-\frac{3g_{L}^{2}}{\sqrt{3g_{L}^{2}-g_{Y}^{2}}}x_{\alpha}\right)\overline{\ell}_{\alpha}\gamma^{\mu}\ell_{\alpha}Z_{\mu}'\,,
\end{align}
with $y_{e,\mu,\tau}=-1/2$, $x_{e}=-1/3$ and $x_{\mu,\tau}=-2/3$.
After rotation to the mass basis, with 
a matrix $V_{l_{\alpha}l_{\beta}}$ ($\alpha$, $\beta$=$e$, $\mu$, $\tau$), we find: 
\begin{align}
\mathscr{L}_{\ell Z'} & =i\left(C_{ee}\overline{e}_{L}\gamma^{\mu}e_{L}+C_{e\mu}\overline{e}_{L}\gamma^{\mu}\mu_{L}\right)Z_{\mu}'\;\textrm{ with }\;C_{ee}\approx\frac{g_{Y}^{2}-g_{L}^{2}}{2\sqrt{3g_{L}^{2}-g_{Y}^{2}}}\;\textrm{ and }\;C_{e\mu}=-\frac{g_{L}^{2}}{\sqrt{3g_{L}^{2}-g_{Y}^{2}}}V_{\ell_{e}\ell_{\mu}}\,,
\end{align}
so \cite{Kuno2001} 
\begin{align}\label{eq:BrMuEEE}
\textrm{Br}\left(\mu\rightarrow e\overline{e}e\right) & =\left|\frac{C_{ee}C_{e\mu}}{G_{F}m_{Z'}^{2}}\right|^{2}\approx8\left(\frac{1-2\sin^{2}\theta_{w}}{3-4\sin^{2}\theta_{w}}\right)^{2}\left|V_{\ell_{e}\ell_{\mu}}\right|^{2}\left(\frac{m_{W}}{m_{Z'}}\right)^{4}\,.
\end{align}

However, recall that $\ell_{e}$ and $\ell_{\mu}$ do not mix at tree
level. After taking into account radiative corrections, one expects 
that not only a non-zero electron mass will appear, but also a non-zero 
mixing element $V_{\ell_{e}\ell_{\mu}}$, of the order of $m_{e}/m_{\mu}$,
will be generated. Thus, we obtain the simple estimate 
\begin{equation}\label{eq:zpp}
\textrm{Br}\left(\mu\rightarrow e\overline{e}e\right) \apprle 
10^{-12}\left(\frac{\textrm{5 TeV}}{m_{Z'}}\right)^{4}
\left|\frac{V_{\ell_{e}\ell_{\mu}}}{(m_{e}/m_{\mu})}\right|^{2}.
\end{equation}
Given that this branching ratio is known to be smaller than
approximately $10^{-12}$ \cite{Bellgardt1988}, it provides a lower
bound on the $Z'$ mass comparable to the LHC one ($m_{Z'}\gtrsim3$
TeV) \cite{ATLAS_Zp,CMS:2015nhc}. Note that equation \eqref{eq:zpp} 
depends quadratically on $V_{\ell_{e}\ell_{\mu}}$.

In models such as this one, the $Z$ boson mass eigenstate has a slightly different composition than in the Standard Model, which is a consequence of the so-called $Z$-$Z'$ mixing. In practice, this means that there is an analogue to the diagram in figure \ref{fig:muToeee} where the $Z'$ is replaced by a $Z$ boson. However, the $Z_\nu \overline{\mu}_{L} \gamma^\nu e_L$ interaction is suppressed by a factor $\rho m_Z^2/m^2_{Z'}$ relative to the $Z'_\nu\overline{\mu}_{L}\gamma^\nu e_L$ interaction. It turns out that $\left|\rho\right|<1$, so at most we expect a $\mathcal{O}\left(1\right)$ correction to equation \eqref{eq:zpp} due to $Z$-$Z'$ mixing.\footnote{Given that the interaction of the $Z$ to electrons is somewhat stronger than the one of $Z'$, even though $\rho$ can never be $-1$, it still seems possible, even if improbable, to have a complete suppression of the $\mu\rightarrow e\overline{e}e$ process. Note that a rigorous analysis of such a delicate cancellation would require also taking into account the decay of the muon to right-handed electrons which, for simplicity, was not included in equation \eqref{eq:BrMuEEE}.}

We mention in passing that there will also be some constraints from other decays 
such as $\ell_{i}\rightarrow\ell_{j}\gamma$, which are induced by loops
with the heavy gauge bosons or the heavy fermions. However, 
we expect  the decay $\mu\rightarrow e\overline{e}e$ to 
give the most stringent bound on the 331 breaking scale.

\section{Summary}

This work describes for the first time a model based on the
$SU(3)_{C}\times SU(3)_{L}\times U(1)_{X}$ gauge group where all three
quark families are placed in equal representations of the gauge
group, while the lepton families are not. The required number of
fermions representations is the same as in a standard 331 model. There
are no gauge anomalies, and the observed fermion masses and mixings
can be accommodated with a minimal scalar sector composed of three
triplets and a sextet. Unlike in standard 331 models where the
parameter $\beta$ controlling the relation $Y=\beta T_{8}+X$ can take
more than one value, in our flipped 331 construction $\beta$ must be
$1/\sqrt{3}$.

With the minimal setup considered in this paper, one charged lepton
(the electron) and two neutrinos are massless at tree level. Radiative
effects can generate the missing masses and mixing angles, favoring
an inverse neutrino mass hierarchy, with $\nu_{3}$ almost massless. If the
scalar sector of the model is extended, other constructions might
be possible.

As in all standard 331 models, there are potentially dangerous flavor
violating $Z'$ interactions. However, unlike in other 331
models, in our flipped scenario there are no constraints from the
quark sector, and $Z'$ interactions in the model cannot explain 
the known hints of anomalies in $K$ and $B$ physics. 
Instead, amplitudes of lepton number violating processes are expected to
be sizable. For example, if the $Z'$ is observed at the LHC, we expect that
the proposed Mu3e experiment \cite{Blondel:2013ia} would very likely observe 
the decay $\mu\rightarrow e\overline{e}e$.

\section*{\vspace{-0.15cm}Acknowledgments}

This work was supported by the Spanish grants FPA2014-58183-P, Multidark
CSD2009-00064 and SEV-2014-0398 (from the \textit{Ministerio de Economía
y Competitividad}), as well as PROMETEOII/2014/084 (from the \textit{Generalitat
Valenciana}).


\begin{thebibliography}{10}
	\providecommand{\url}[1]{\texttt{#1}}
	\providecommand{\urlprefix}{URL }
	\providecommand{\eprint}[2][]{\url{#2}}
	
	\bibitem{Schechter:1974dy}
	J.~Schechter and Y.~Ueda, \emph{{Unified weak-electromagnetic gauge schemes
			based on the three-dimensional unitary group}},
	\MYhref[journalLinks]{http://dx.doi.org/10.1103/PhysRevD.8.484}{Phys. Rev.
	}\MYhref[journalLinks]{http://dx.doi.org/10.1103/PhysRevD.8.484}{\textbf{D8}
	(1973) 484--490}.

\bibitem{Gupta:1973pv}
V.~Gupta and H.~S. Mani, \emph{{Unified weak and electromagnetic gauge theory
		based on $SU(3) \times U(1)$}},
\MYhref[journalLinks]{http://dx.doi.org/10.1103/PhysRevD.10.1310}{Phys. Rev.
}\MYhref[journalLinks]{http://dx.doi.org/10.1103/PhysRevD.10.1310}{\textbf{D10}
(1974) 1310}.

\bibitem{Albright:1974nd}
C.~H. Albright, C.~Jarlskog and M.~O. Tjia, \emph{{Implications of gauge
		theories for heavy leptons}},
\MYhref[journalLinks]{http://dx.doi.org/10.1016/0550-3213(75)90360-0}{Nucl.
	Phys.
}\MYhref[journalLinks]{http://dx.doi.org/10.1016/0550-3213(75)90360-0}{\textbf{B86}
(1975) 535--547}.

\bibitem{Georgi:1978bv}
H.~Georgi and A.~Pais, \emph{{Generalization of the Glashow-Iliopoulos-Maiani
		mechanism: horizontal and vertical flavor mixing}},
\MYhref[journalLinks]{http://dx.doi.org/10.1103/PhysRevD.19.2746}{Phys. Rev.
}\MYhref[journalLinks]{http://dx.doi.org/10.1103/PhysRevD.19.2746}{\textbf{D19}
(1979) 2746}.

\bibitem{Singer1980}
M.~Singer, J.~Valle and J.~Schechter, \emph{{Canonical neutral-current
		predictions from the weak-electromagnetic gauge group $SU(3) \times U(1) $}},
\MYhref[journalLinks]{http://dx.doi.org/10.1103/PhysRevD.22.738}{Phys. Rev. D
}\MYhref[journalLinks]{http://dx.doi.org/10.1103/PhysRevD.22.738}{\textbf{22}
(1980) 738}.

\bibitem{Pisano1992}
F.~Pisano and V.~Pleitez, \emph{{$SU(3)\times U(1)$ model for electroweak
		interactions}},
\MYhref[journalLinks]{http://dx.doi.org/10.1103/PhysRevD.46.410}{Phys. Rev. D
}\MYhref[journalLinks]{http://dx.doi.org/10.1103/PhysRevD.46.410}{\textbf{46}
(1992) 410--417},
\MYhref[eprintLinks]{http://arxiv.org/abs/hep-ph/9206242}{{\ttfamily
		arXiv:hep-ph/9206242}}.

\bibitem{Frampton1992}
P.~H. Frampton, \emph{{Chiral dilepton model and the flavor question}},
\MYhref[journalLinks]{http://dx.doi.org/10.1103/PhysRevLett.69.2889}{Phys.
	Rev. Lett.
}\MYhref[journalLinks]{http://dx.doi.org/10.1103/PhysRevLett.69.2889}{\textbf{69}
(1992) 2889--2891}.

\bibitem{Diaz2005}
R.~A. Diaz, R.~Martinez and F.~Ochoa, \emph{{$SU(3)_C \times SU(3)_L \times
		U(1)_X$ models for $\beta$ arbitrary and families with mirror fermions}},
\MYhref[journalLinks]{http://dx.doi.org/10.1103/PhysRevD.72.035018}{Phys.
	Rev.
}\MYhref[journalLinks]{http://dx.doi.org/10.1103/PhysRevD.72.035018}{\textbf{D72}
(2005) 035018},
\MYhref[eprintLinks]{http://arxiv.org/abs/hep-ph/0411263}{{\ttfamily
		arXiv:hep-ph/0411263 [hep-ph]}}.

\bibitem{Buras:2014yna}
A.~J. Buras, F.~De~Fazio and J.~Girrbach-Noe, \emph{{Z-Z' mixing and Z-mediated
		FCNCs in $SU(3)_{C} \times SU(3)_{L} \times U(1)_{X}$ models}},
\MYhref[journalLinks]{http://dx.doi.org/10.1007/JHEP08(2014)039}{JHEP
}\MYhref[journalLinks]{http://dx.doi.org/10.1007/JHEP08(2014)039}{\textbf{08}
(2014) 039}, \MYhref[eprintLinks]{http://arxiv.org/abs/1405.3850}{{\ttfamily
	arXiv:1405.3850 [hep-ph]}}.

\bibitem{Foot1993}
R.~Foot, O.~F. Hern\'andez, F.~Pisano and V.~Pleitez, \emph{{Lepton masses in an
		$SU(3)_L \times U(1)_N$ gauge model}},
\MYhref[journalLinks]{http://dx.doi.org/10.1103/PhysRevD.47.4158}{Phys. Rev.
}\MYhref[journalLinks]{http://dx.doi.org/10.1103/PhysRevD.47.4158}{\textbf{D47}
(1993) 4158--4161},
\MYhref[eprintLinks]{http://arxiv.org/abs/hep-ph/9207264}{{\ttfamily
		arXiv:hep-ph/9207264 [hep-ph]}}.

\bibitem{Fonseca2016}
R.~M. Fonseca and M.~Hirsch, \emph{{Lepton number violation in 331 models}}
(2016), \MYhref[eprintLinks]{http://arxiv.org/abs/1607.06328}{{\ttfamily
		 arXiv:1607.06328 [hep-ph]}}.

\bibitem{Sanchez2001}
L.~A. Sanchez, W.~A. Ponce and R.~Martinez, \emph{{$SU(3)_c \times SU(3)_L
		\times U(1)_X$ as an $E(6)$ subgroup}},
\MYhref[journalLinks]{http://dx.doi.org/10.1103/PhysRevD.64.075013}{Phys.
	Rev.
}\MYhref[journalLinks]{http://dx.doi.org/10.1103/PhysRevD.64.075013}{\textbf{D64}
(2001) 075013},
\MYhref[eprintLinks]{http://arxiv.org/abs/hep-ph/0103244}{{\ttfamily
		arXiv:hep-ph/0103244 [hep-ph]}}.

\bibitem{Fonseca2015}
R.~M. Fonseca, \emph{{On the chirality of the SM and the fermion content of
		GUTs}},
\MYhref[journalLinks]{http://dx.doi.org/10.1016/j.nuclphysb.2015.06.012}{Nucl.
	Phys.
}\MYhref[journalLinks]{http://dx.doi.org/10.1016/j.nuclphysb.2015.06.012}{\textbf{B897}
(2015) 757--780},
\MYhref[eprintLinks]{http://arxiv.org/abs/1504.03695}{{\ttfamily
		arXiv:1504.03695 [hep-ph]}}.

\bibitem{Forero2014}
D.~V. Forero, M.~T\'ortola and J.~W.~F. Valle, \emph{{Neutrino oscillations
		refitted}},
\MYhref[journalLinks]{http://dx.doi.org/10.1103/PhysRevD.90.093006}{Phys.
	Rev.
}\MYhref[journalLinks]{http://dx.doi.org/10.1103/PhysRevD.90.093006}{\textbf{D90}
(2014) 9 093006},
\MYhref[eprintLinks]{http://arxiv.org/abs/1405.7540}{{\ttfamily
		arXiv:1405.7540 [hep-ph]}}.

\bibitem{Buras:2015kwd}
A.~J. Buras and F.~De~Fazio, \emph{{$\varepsilon'/\varepsilon$ in 331 models}},
\MYhref[journalLinks]{http://dx.doi.org/10.1007/JHEP03(2016)010}{JHEP
}\MYhref[journalLinks]{http://dx.doi.org/10.1007/JHEP03(2016)010}{\textbf{03}
(2016) 010}, \MYhref[eprintLinks]{http://arxiv.org/abs/1512.02869}{{\ttfamily
	arXiv:1512.02869 [hep-ph]}}.

\bibitem{Boucenna2015b}
S.~M. Boucenna, J.~W.~F. Valle and A.~Vicente, \emph{{Predicting charged lepton
		flavor violation from 3-3-1 gauge symmetry}},
\MYhref[journalLinks]{http://dx.doi.org/10.1103/PhysRevD.92.053001}{Phys.
	Rev.
}\MYhref[journalLinks]{http://dx.doi.org/10.1103/PhysRevD.92.053001}{\textbf{D92}
(2015) 5 053001},
\MYhref[eprintLinks]{http://arxiv.org/abs/1502.07546}{{\ttfamily
		arXiv:1502.07546 [hep-ph]}}.

\bibitem{Kuno2001}
Y.~Kuno and Y.~Okada, \emph{{Muon decay and physics beyond the standard
		model}},
\MYhref[journalLinks]{http://dx.doi.org/10.1103/RevModPhys.73.151}{Rev. Mod.
	Phys.
}\MYhref[journalLinks]{http://dx.doi.org/10.1103/RevModPhys.73.151}{\textbf{73}
(2001) 151--202},
\MYhref[eprintLinks]{http://arxiv.org/abs/hep-ph/9909265}{{\ttfamily
		arXiv:hep-ph/9909265 [hep-ph]}}.

\bibitem{Bellgardt1988}
U.~Bellgardt et~al. (SINDRUM), \emph{{Search for the decay $\mu^{+}\rightarrow
		e^{+}e^{+}e^{-}$}},
\MYhref[journalLinks]{http://dx.doi.org/10.1016/0550-3213(88)90462-2}{Nucl.
	Phys.
}\MYhref[journalLinks]{http://dx.doi.org/10.1016/0550-3213(88)90462-2}{\textbf{B299}
(1988) 1--6}.

\bibitem{ATLAS_Zp}
ATLAS collaboration, \emph{{Search for new phenomena in the dilepton final
		state using proton-proton collisions at $\sqrt{s}=$ 13 TeV with the ATLAS
		detector}}, \MYhref[eprintLinks]{http://cds.cern.ch/record/2114842}{{\ttfamily
		ATLAS-CONF-2015-070}}  (2015).

\bibitem{CMS:2015nhc}
CMS~collaboration, \emph{{Search for a narrow resonance produced in 13 TeV
		pp collisions decaying to electron pair or muon pair final states}}, \MYhref[eprintLinks]{https://cds.cern.ch/record/2114855}{{\ttfamily
		CMS-PAS-EXO-15-005}} (2015).

\bibitem{Blondel:2013ia}
A.~Blondel et~al., \emph{{Research proposal for an experiment to search for the
		decay $\mu \to eee$}}  (2013),
\MYhref[eprintLinks]{http://arxiv.org/abs/1301.6113}{{\ttfamily
		arXiv:1301.6113 [physics.ins-det]}}.

\end{thebibliography}
\end{document}